\documentclass[journal]{IEEEtran}
\usepackage{cite}

\ifCLASSINFOpdf
  \usepackage[pdftex]{graphicx}
  \DeclareGraphicsExtensions{.pdf,.jpeg,.png}
\else
  \usepackage[dvips]{graphicx}
  \DeclareGraphicsExtensions{.eps}
\fi
\usepackage{amsmath}
\begin{document}
%
\title{Fast analysis of spherical metasurfaces using vector wave function expansion}

%
%
%

\author{Srikumar~Sandeep,
        Shao~Ying~Huang,~\IEEEmembership{Member,~IEEE}
\thanks{S.\ Sandeep (email: sandeepsrikumar2013@gmail.com) and S. Y. \ Huang (email: huangshaoying@sutd.edu.sg) are with Singapore University of Technology and Design, Singapore}
}
%

%
%

\markboth{IEEE ANTENNAS AND WIRELESS PROPAGATION LETTERS}
{Shell \MakeLowercase{\textit{et al.}}: Bare Demo of IEEEtran.cls for IEEE Journals}
%



\maketitle

\begin{abstract}
Modeling of spherical metasurfaces using Generalized Sheet Transition Conditions (GSTCs) and Vector Wave Function (VWF) expansion is presented. The fields internal and external to the metasurface is expanded in terms of spherical VWFs and unknown coefficients. GSTCs are used to obtain linear relationships between the unknown coefficients. An overdetermined system of equations are then solved by point matching. The method is quasi-analytical and hence there is no need for meshing which is encountered in conventional computational electromagnetic methods. This results in the method being extremely fast and hence useful for metasurface optimization. The method is validated by two examples. The formulations presented here can be easily extended to multilayered spherical metasurfaces. 
\end{abstract}


%
\IEEEpeerreviewmaketitle

\section{Introduction}
%
%
%
%
\IEEEPARstart{M}{etasurfaces} are deeply subwavelength surfaces which can manipulate electromagnetic waves in a desired manner \cite{HollawayMetasurfaceReview}. Essentially, these are field transformers which are constructed by arrangement of subwavelength scatterers in a host medium. Metasurfaces have practical advantages over bulk metamaterials including easier fabrication, lower loss and less weight \cite{MetamaterialsSolymar}. Even though they have similarities with frequency selective surfaces, the design possibilities offered by metasurfaces are much broader. Metasurface applications include generalized refraction \cite{NYuLightProp}, flat optical components \cite{FlatOpticsYu}, LED efficiency enhancers \cite{LEDEnhancerMS}, spatial isolators \cite{SpatialIsolatorMS} etc. Review of metasurface and its applications can be found in \cite{RecentDevelopmentsMSAchouri,MetasurfaceReviewGlybovski}.

Metasurfaces achieve its functionality by creating a spatio-temporal electromagnetic discontinuity. Mathematically, the discontinuity can be expressed by Generalized Sheet Transition Conditions (GSTC) which relates the electric and magnetic field discontinuities to the electric and magnetic surface polarization current densities \cite{IdemenBook,achouri2015general}.  At present, commercial electromagnetic simulation softwares can model several boundary conditions, such as perfect electric conductor (PEC), perfect magnetic conductor (PMC), periodic boundary condition (PBC), standard impedance boundary condition (SIBC), radiation boundary condition (RBC), and perfectly matched layer (PML). However, no commercial CAD tools have yet incorporated the modeling of GSTCs. Therefore, it is important to develop numerical modeling of GSTCs for analysis and synthesis of metasurfaces. The modeling of GSTCs in the Finite Difference Frequency Domain (FDFD) method was reported in \cite{GSTCFDFD}. Modeling of GSTCs in the Finite Element Method (FEM), which is one of the more widely used numerical methods to simulate practical problems was described in \cite{GSTCFEM}. Recently, Integral Equation - Method of Moment (IE-MoM) was applied to circular cylindrical metasurfaces \cite{CircCylMSMoM} and cylindrical metasurfaces of arbitrary cross-section \cite{CylMSMoM}. A review of computational electromagnetic methods applied to metasurface analysis can be found in \cite{ComputationalAnalysisMS}.

The vast majority of metasurfaces reported to date are planar. Other canonical shapes such as cylindrical metasurfaces and spherical metasurfaces \cite{SphericalMS} are now being studied. It is expected that conformal metasurfaces (metasurfaces of irregular shape) will become a subject of active research \cite{SphericalMS}. Some of the applications of spherical metasurfaces include mantle cloaking\cite{MantleCloakingMTS}  and radiation pattern control \cite{SphericalMTSRadPatControl}. In this paper, fast analysis of spherical metasurfaces is performed using vector wave functon expansion. Since this method does not involve meshing, it is considerably faster than conventional computational electromagnetic methods such as MoM. It should be noted that physical metasurfaces have a finite subwavelength thickness. Simulating such structures directly would result in very dense meshes around the metasurfaces and hence compromise the simulation efficiency. By replacing a physical metasurface by an equivalent GSTC, the burden of mesh generation can be reduced significantly and the simulation efficiency can be enhanced considerably. This is particularly important in simulation scenarios where multiple metasurfaces are involved or when repetitive simulations are required for physical metasurface design and optimization \cite{achouri2015general,ComputationalAnalysisMS}.  

The organization of the paper is as follows. Section II recalls the GSTC metasurface synthesis equations for spherical metasurfaces. This is followed by the application of spherical VWFs combined with GSTCs to the metasurface scattering problem in section III. Section IV contains two validation examples. Conclusions are provided in Section V. 


\section{Spherical metasurface scattering problem}
Metasurface synthesis equations for planar metasurfaces and spherical metasurfaces are described in \cite{achouri2015general} and \cite{SphericalMS} respectively. Similar to \cite{achouri2015general} and \cite{SphericalMS}, the normal polarization current densities are ignored are this work.  Following the bianisotropic susceptibility-GSTC approach \cite{SphericalMS}, the metasurface synthesis equations for a spherical metasurface of radius $a$  reads
\begin{subequations}
\label{SphericalMSSynthesisEqns}
\begin{equation} \label{GSTCexpandedeqns1}
\begin{split}
\begin{bmatrix}
-\Delta H_{\phi} \\
\Delta H_{\theta} 
\end{bmatrix}
 =  j\omega\epsilon_{0}
\begin{bmatrix}
\chi_{\mathrm{ee}}^{\theta \theta} & \chi_{\mathrm{ee}}^{\theta \phi} \\
\chi_{\mathrm{ee}}^{\phi \theta} & \chi_{\mathrm{ee}}^{\phi \phi}
\end{bmatrix}
\begin{bmatrix}
E_{\theta,\mathrm{av}} \\
E_{\phi,\mathrm{av}}
\end{bmatrix}
  + j\omega\sqrt{\mu_{0} \epsilon_{0}} \\
\begin{bmatrix}
\chi_{\mathrm{em}}^{\theta \theta} & \chi_{\mathrm{em}}^{\theta \phi} \\
\chi_{\mathrm{em}}^{\phi \theta} & \chi_{\mathrm{em}}^{\phi \phi}
\end{bmatrix}
\begin{bmatrix}
H_{\theta,\mathrm{av}} \\
H_{\phi,\mathrm{av}}
\end{bmatrix} 
\end{split}
\end{equation} 
\begin{equation}
\begin{split}
\label{GSTCexpandedeqns2}
\begin{bmatrix}
\Delta E_{\phi} \\
-\Delta E_{\theta}
\end{bmatrix}
  =  j\omega \mu_{0}
\begin{bmatrix}
\chi_{\mathrm{mm}}^{\theta \theta} & \chi_{\mathrm{mm}}^{\theta \phi} \\
\chi_{\mathrm{mm}}^{\phi \theta} & \chi_{\mathrm{mm}}^{\phi \phi}
\end{bmatrix}
\begin{bmatrix}
H_{\theta,\mathrm{av}} \\
H_{\phi,\mathrm{av}}
\end{bmatrix}
  + j\omega \sqrt{\mu_{0} \epsilon_{0}} \\
\begin{bmatrix}
\chi_{\mathrm{me}}^{\theta \theta} & \chi_{\mathrm{me}}^{\theta \phi} \\
\chi_{\mathrm{me}}^{\phi \theta} & \chi_{\mathrm{me}}^{\phi \phi}
\end{bmatrix}
\begin{bmatrix}
E_{\theta,\mathrm{av}} \\
E_{\phi,\mathrm{av}}
\end{bmatrix}
\end{split}
\end{equation}
\end{subequations}
where $\overline{\overline{\chi}}_{\mathrm{ee}},\overline{\overline{\chi}}_{\mathrm{mm}},\overline{\overline{\chi}}_{\mathrm{em}}$, and $\overline{\overline{\chi}}_{\mathrm{me}}$ are the electric/magnetic (first e/m subscripts) susceptibility tensors describing the response to the electric/magnetic (second e/m subscripts) excitations. $\Delta \Psi = \Psi^{+} - \Psi^{-}$ denote the jump discontinuity of field component $\Psi$ and the subscript ``av'' denotes the average of the fields on both sides of the metasurface, $\vec{\psi}_{\mathrm{av}} = [(\vec{\psi}^{\mathrm{inc}} + \vec{\psi}^{\mathrm{ref}}) + \vec{\psi}^{\mathrm{tr}}]/2$. The medium internal and external to the metasurface sphere is free space and a time harmonic dependance of $e^{j\omega t}$ is assumed. Through out this work, we have assumed a monoanisotropic metasurface, i.e. $\overline{\overline{\chi}}_{\mathrm{em}} = \overline{\overline{\chi}}_{\mathrm{me}} = 0$. In such a case, the metasurface synthesis equations simplifies to 
\begin{subequations}
\label{SphMTSSynthesisEqnsMonoAniso}
\begin{equation} \label{GSTCexpandedeqns1MonoAniso}
\begin{split}
\begin{bmatrix}
-\Delta H_{\phi} \\
\Delta H_{\theta} 
\end{bmatrix}
 =  j\omega\epsilon_{0}
\begin{bmatrix}
\chi_{\mathrm{ee}}^{\theta \theta} & \chi_{\mathrm{ee}}^{\theta \phi} \\
\chi_{\mathrm{ee}}^{\phi \theta} & \chi_{\mathrm{ee}}^{\phi \phi}
\end{bmatrix}
\begin{bmatrix}
E_{\theta,\mathrm{av}} \\
E_{\phi,\mathrm{av}}
\end{bmatrix} 
\end{split}
\end{equation} 
\begin{equation}
\begin{split}
\label{GSTCexpandedeqns2MonoAniso}
\begin{bmatrix}
\Delta E_{\phi} \\
-\Delta E_{\theta}
\end{bmatrix}
  =  
 j\omega \mu_{0} 
\begin{bmatrix}
\chi_{\mathrm{mm}}^{\theta \theta} & \chi_{\mathrm{mm}}^{\theta \phi} \\
\chi_{\mathrm{mm}}^{\phi \theta} & \chi_{\mathrm{mm}}^{\phi \phi}
\end{bmatrix}
\begin{bmatrix}
H_{\theta,\mathrm{av}} \\
H_{\phi,\mathrm{av}}
\end{bmatrix}
\end{split}
\end{equation}
\end{subequations}

\begin{figure}[!h]
\centering
\includegraphics[width=2in]{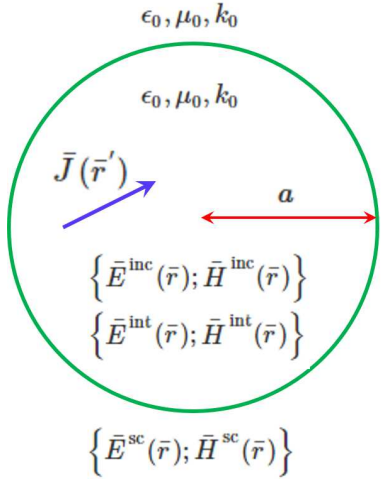}
\caption{Spherical metasurface scattering problem.}
\label{fig_SphericalMTSProb}
\end{figure}
The spherical metasurface scattering problem is depicted in Fig. \ref{fig_SphericalMTSProb}. Given a spherical metasurface with  known susceptibility tensors and an arbitrary electric current density source internal to the metasurface, the problem is to find the total field internal and external to the metasurface. Inspired by Mie scattering \cite{BohrenHuffman}, the field inside the spherical metasurface can be denoted as a sum of known incident field ($\bar{E}^{\mathrm{inc}};\bar{H}^{\mathrm{inc}}$) and unknown internal field ($\bar{E}^{\mathrm{int}};\bar{H}^{\mathrm{int}}$). The field external to the metasurface is denoted by scattered field ($\bar{E}^{\mathrm{sc}};\bar{H}^{\mathrm{sc}}$). The incident field is generated by the source electric current density $\bar{J}(\bar{r}^{'})$ in a homogenous medium (i.e. no metasurface). Given a $\bar{J}(\bar{r}^{'})$, the incident fields can be easily obtained using either free space dyadic green function or vector potential radiation integrals. Thus the spherical metasurface scattering problem reduces to finding the internal and the scattered field. It should be mentioned that internal field is just the scattered field internal to the metasurface. It is referred to as internal to differentiate it from the external scattered field. The application of these three field sets in (\ref{SphMTSSynthesisEqnsMonoAniso}) results in
\begin{subequations}
\label{SphMTSEqnVWF}
\begin{equation} \label{SphMTSEqnVWF1}
\begin{split}
\begin{bmatrix}
-H_{\phi}^{\mathrm{sc}} +  H_{\phi}^{\mathrm{int}}\\
H_{\theta}^{\mathrm{sc}} -  H_{\theta}^{\mathrm{int}}
\end{bmatrix}
 -  \frac{j\omega\epsilon_{0}}{2}
\begin{bmatrix}
\chi_{\mathrm{ee}}^{\theta \theta} & \chi_{\mathrm{ee}}^{\theta \phi} \\
\chi_{\mathrm{ee}}^{\phi \theta} & \chi_{\mathrm{ee}}^{\phi \phi}
\end{bmatrix}
\begin{bmatrix}
E_{\theta}^{\mathrm{sc}} + E_{\theta}^{\mathrm{int}} \\
E_{\phi}^{\mathrm{sc}} + E_{\phi}^{\mathrm{int}}  
\end{bmatrix} 
\\
=
\begin{bmatrix}
-H_{\phi}^{\mathrm{inc}} \\
H_{\theta}^{\mathrm{inc}} 
\end{bmatrix}
 +  \frac{j\omega\epsilon_{0}}{2}
\begin{bmatrix}
\chi_{\mathrm{ee}}^{\theta \theta} & \chi_{\mathrm{ee}}^{\theta \phi} \\
\chi_{\mathrm{ee}}^{\phi \theta} & \chi_{\mathrm{ee}}^{\phi \phi}
\end{bmatrix}
\begin{bmatrix}
E_{\theta}^{\mathrm{inc}}  \\
E_{\phi}^{\mathrm{inc}} 
\end{bmatrix} 
\end{split}
\end{equation} 
\begin{equation} \label{SphMTSEqnVWF2}
\begin{split}
\begin{bmatrix}
E_{\phi}^{\mathrm{sc}} -  E_{\phi}^{\mathrm{int}} \\
-E_{\theta}^{\mathrm{sc}} +  E_{\theta}^{\mathrm{int}}
\end{bmatrix}
 -  \frac{j\omega\mu_{0}}{2}
\begin{bmatrix}
\chi_{\mathrm{mm}}^{\theta \theta} & \chi_{\mathrm{mm}}^{\theta \phi} \\
\chi_{\mathrm{mm}}^{\phi \theta} & \chi_{\mathrm{mm}}^{\phi \phi}
\end{bmatrix}
\begin{bmatrix}
H_{\theta}^{\mathrm{sc}} + H_{\theta}^{\mathrm{int}} \\
H_{\phi}^{\mathrm{sc}} + H_{\phi}^{\mathrm{int}}  
\end{bmatrix} 
\\
=
\begin{bmatrix}
E_{\phi}^{\mathrm{inc}} \\
-E_{\theta}^{\mathrm{inc}} 
\end{bmatrix}
 +  \frac{j\omega\mu_{0}}{2}
\begin{bmatrix}
\chi_{\mathrm{mm}}^{\theta \theta} & \chi_{\mathrm{mm}}^{\theta \phi} \\
\chi_{\mathrm{mm}}^{\phi \theta} & \chi_{\mathrm{mm}}^{\phi \phi}
\end{bmatrix}
\begin{bmatrix}
H_{\theta}^{\mathrm{inc}}  \\
H_{\phi}^{\mathrm{inc}} 
\end{bmatrix} 
\end{split}
\end{equation}
\end{subequations}
In the above four equations, the left hand side contains the unknown field variables and the right hand side contains the known incident field. 

\section{Spherical vector wave function expansion}
Vector Wave Functions (VWF) are eigen functions of vector wave equation \cite{BohrenHuffman}. Spherical vector wave functions are given by 
\begin{equation}
\begin{split}
\bar{M}^{(i)}_{\begin{subarray}{l}e\\o \end{subarray}nm}(\bar{r};k_{0})
= \mp mz_{n}^{(i)}(k_{0}r) \frac{P_{n}^{m}(\cos \theta)}{\sin \theta} 
\bigg\{ 
\!
\begin{array}{l}
\sin m \phi 
\\
\cos m \phi
\end{array} \!
\bigg\}
\hat{\theta} \\
- z_{n}^{(i)}(k_{0}r) \frac{dP_{n}^{m}(\cos \theta)}{d \theta} 
\bigg\{ 
\begin{array}{l}
\cos m \phi 
\\
\sin m \phi
\end{array}
\bigg\} 
\hat{\phi}
\end{split}
\end{equation}
\begin{equation}
\begin{split}
\bar{N}^{(i)}_{\begin{subarray}{l}e\\o \end{subarray}nm}(\bar{r};k_{0})
= \frac{n(n+1)z_{n}^{(i)}(k_{0}r)}{k_{0}r} P_{n}^{m}(\cos \theta)
\bigg\{ 
\!
\begin{array}{l}
\cos m \phi 
\\
\sin m \phi
\end{array} \!
\bigg\}
\hat{r} \\
+ \frac{1}{k_{0}r}\frac{d}{dr}(rz_{n}^{(i)}(k_{0}r)) \frac{dP_{n}^{m}(\cos \theta)}{d \theta} 
\bigg\{ 
\begin{array}{l}
\cos m \phi 
\\
\sin m \phi
\end{array}
\bigg\} 
\hat{\theta} \\
\mp \frac{m}{k_{0}r}\frac{d}{dr}(rz_{n}^{(i)}(k_{0}r)) \frac{P_{n}^{m}(\cos \theta)}{\sin \theta} 
\bigg\{ 
\begin{array}{l}
\sin m \phi 
\\
\cos m \phi
\end{array}
\bigg\} 
\hat{\phi}
\end{split}
\end{equation}
where $z_{n}^{(i)}(k_{0}r)$ can be either one of the four spherical Bessel / Hankel functions of first / second kind. $z_{n}^{(i)}(k_{0}r) = \{j_{n}(k_{0}r),y_{n}(k_{0}r),h_{n}^{(1)}(k_{0}r),h_{n}^{(2)}(k_{0}r)\}$ for $i = 1,2,3,4$ respectively. $P_{n}^{m}(\cos \theta)$ is the associated Legendre function of order $m$ and degree $n$. $k_{0}$ is the wave number. 
The unknown electric field vectors $\bar{E}^{\mathrm{sc}}(\bar{r})$ and $\bar{E}^{\mathrm{int}}(\bar{r})$ are expressed as an eigen function expansion in terms of unknowns coefficients and known spherical vector wave functions. 
\begin{equation}
\label{EscExp}
\bar{E}^{\mathrm{sc}}(\bar{r}) \approx \sum_{n=1}^{N} \sum_{m=0}^{n} A_{\begin{subarray}{l}e\\o \end{subarray}nm}
\bar{M}^{(4)}_{\begin{subarray}{l}e\\o \end{subarray}nm}(\bar{r};k_{0}) + 
B_{\begin{subarray}{l}e\\o \end{subarray}nm}
\bar{N}^{(4)}_{\begin{subarray}{l}e\\o \end{subarray}nm}(\bar{r};k_{0})
\end{equation}
\begin{equation}
\label{EintExp}
\bar{E}^{\mathrm{int}}(\bar{r}) \approx \sum_{n=1}^{N} \sum_{m=0}^{n} C_{\begin{subarray}{l}e\\o \end{subarray}nm}
\bar{M}^{(1)}_{\begin{subarray}{l}e\\o \end{subarray}nm}(\bar{r};k_{0}) + 
D_{\begin{subarray}{l}e\\o \end{subarray}nm}
\bar{N}^{(1)}_{\begin{subarray}{l}e\\o \end{subarray}nm}(\bar{r};k_{0})
\end{equation}
The summation in $n$ is truncated to $N$ for numerical implementation and hence approximate symbol is used in the above equations. Since the internal region contains the origin and spherical Neumann function is singular at the origin, only the spherical Bessel function of first kind is used in the internal field expression.The magnetic fields $\bar{H}^{\mathrm{sc}}(\bar{r}), \bar{H}^{\mathrm{int}}(\bar{r})$ are obtained by applying Maxwell-Faraday equation on $(\ref{EscExp}), (\ref{EintExp})$.
\begin{equation}
\label{HscExp}
\bar{H}^{\mathrm{sc}}(\bar{r}) \approx \frac{j}{\eta_{0}}\sum_{n=1}^{N} \sum_{m=0}^{n} A_{\begin{subarray}{l}e\\o \end{subarray}nm}
\bar{N}^{(4)}_{\begin{subarray}{l}e\\o \end{subarray}nm}(\bar{r};k_{0}) + 
B_{\begin{subarray}{l}e\\o \end{subarray}nm}
\bar{M}^{(4)}_{\begin{subarray}{l}e\\o \end{subarray}nm}(\bar{r};k_{0})
\end{equation}
\begin{equation}
\label{HintExp}
\bar{H}^{\mathrm{int}}(\bar{r}) \approx \frac{j}{\eta_{0}}\sum_{n=1}^{N} \sum_{m=0}^{n} C_{\begin{subarray}{l}e\\o \end{subarray}nm}
\bar{N}^{(1)}_{\begin{subarray}{l}e\\o \end{subarray}nm}(\bar{r};k_{0}) + 
D_{\begin{subarray}{l}e\\o \end{subarray}nm}
\bar{M}^{(1)}_{\begin{subarray}{l}e\\o \end{subarray}nm}(\bar{r};k_{0})
\end{equation}
where we have used used the following relationship between the vector wave functions.
\begin{subequations}
\label{Idents}
\begin{equation} \label{Ident1}
\nabla \times \bar{M}^{(i)}_{\begin{subarray}{l}e\\o \end{subarray}nm}(\bar{r};k_{0})
= k_{0} \bar{N}^{(i)}_{\begin{subarray}{l}e\\o \end{subarray}nm}(\bar{r};k_{0})
\end{equation} 
\begin{equation} \label{Ident2}
\nabla \times \bar{N}^{(i)}_{\begin{subarray}{l}e\\o \end{subarray}nm}(\bar{r};k_{0})
= k_{0} \bar{M}^{(i)}_{\begin{subarray}{l}e\\o \end{subarray}nm}(\bar{r};k_{0})
\end{equation}
\end{subequations}
The VWF expansions  (\ref{EscExp}),(\ref{EintExp}),(\ref{HscExp}) and (\ref{HintExp}) are substituted in 
(\ref{SphMTSEqnVWF}) resulting in 4 linear equations in terms of the unknowns coefficients $A_{enm}, A_{onm}, B_{enm}, B_{onm}, C_{enm}, C_{onm}, D_{enm}, D_{onm}$. 
\begin{equation}
\begin{split}
\sum_{n=1}^{N} \sum_{m=0}^{n} 
\\
\left[\frac{j}{\eta_{0}}N^{(4)}_{\begin{subarray}{l}e\\o \end{subarray}nm\theta}-\frac{j\omega \epsilon_{0}}{2}\chi_{\mathrm{ee}}^{\phi \phi}M^{(4)}_{\begin{subarray}{l}e\\o \end{subarray}nm\phi}  - \frac{j\omega \epsilon_{0}}{2}\chi_{\mathrm{ee}}^{\phi \theta}M^{(4)}_{\begin{subarray}{l}e\\o \end{subarray}nm\theta}  \right] A_{\begin{subarray}{l}e\\o \end{subarray}nm} +  
\\
\left[\frac{j}{\eta_{0}}M^{(4)}_{\begin{subarray}{l}e\\o \end{subarray}nm\theta}-\frac{j\omega \epsilon_{0}}{2}\chi_{\mathrm{ee}}^{\phi \phi}N^{(4)}_{\begin{subarray}{l}e\\o \end{subarray}nm\phi} -\frac{j\omega \epsilon_{0}}{2}\chi_{\mathrm{ee}}^{\phi \theta}N^{(4)}_{\begin{subarray}{l}e\\o \end{subarray}nm\theta} \right] B_{\begin{subarray}{l}e\\o \end{subarray}nm} + 
\\ 
\left[-\frac{j}{\eta_{0}}N^{(1)}_{\begin{subarray}{l}e\\o \end{subarray}nm \theta}-\frac{j\omega \epsilon_{0}}{2}\chi_{\mathrm{ee}}^{\phi \phi}M^{(1)}_{\begin{subarray}{l}e\\o \end{subarray}nm \phi} -\frac{j\omega \epsilon_{0}}{2}\chi_{\mathrm{ee}}^{\phi \theta}M^{(1)}_{\begin{subarray}{l}e\\o \end{subarray}nm \theta} \right] C_{\begin{subarray}{l}e\\o \end{subarray}nm} + 
\\
\left[-\frac{j}{\eta_{0}}M^{(1)}_{\begin{subarray}{l}e\\o \end{subarray}nm\theta}-\frac{j\omega \epsilon_{0}}{2}\chi_{\mathrm{ee}}^{\phi \phi}N^{(1)}_{\begin{subarray}{l}e\\o \end{subarray}nm \phi} -\frac{j\omega \epsilon_{0}}{2}\chi_{\mathrm{ee}}^{\phi \theta}N^{(1)}_{\begin{subarray}{l}e\\o \end{subarray}nm \theta}\right] D_{\begin{subarray}{l}e\\o \end{subarray}nm} 
\\
= 
H_{\theta}^{\mathrm{inc}} + \frac{j\omega \epsilon_{0}}{2}\chi_{ee}^{\phi \phi}E_{\phi}^{\mathrm{inc}}
 + \frac{j\omega \epsilon_{0}}{2}\chi_{ee}^{\phi \theta}E_{\theta}^{\mathrm{inc}}
\end{split}
\end{equation}
\begin{equation}
\begin{split}
\sum_{n=1}^{N} \sum_{m=0}^{n} 
\\
 \left[\frac{j}{\eta_{0}}N^{(4)}_{\begin{subarray}{l}e\\o \end{subarray}nm \phi}+\frac{j\omega \epsilon_{0}}{2}\chi_{\mathrm{ee}}^{\theta\theta}M^{(4)}_{\begin{subarray}{l}e\\o \end{subarray}nm \theta} +\frac{j\omega \epsilon_{0}}{2}\chi_{\mathrm{ee}}^{\theta\phi}M^{(4)}_{\begin{subarray}{l}e\\o \end{subarray}nm \phi} \right] A_{\begin{subarray}{l}e\\o \end{subarray}nm } + 
\\
\left[\frac{j}{\eta_{0}}M^{(4)}_{\begin{subarray}{l}e\\o \end{subarray}nm  \phi}+\frac{j\omega \epsilon_{0}}{2}\chi_{\mathrm{ee}}^{\theta \theta}N^{(4)}_{\begin{subarray}{l}e\\o \end{subarray}nm  \theta} 
+\frac{j\omega \epsilon_{0}}{2}\chi_{\mathrm{ee}}^{\theta \phi}N^{(4)}_{\begin{subarray}{l}e\\o \end{subarray}nm  \phi} 
\right] B_{\begin{subarray}{l}e\\o \end{subarray}nm } + 
\\  
\left[-\frac{j}{\eta_{0}}N^{(1)}_{\begin{subarray}{l}e\\o \end{subarray}nm \phi}+\frac{j\omega \epsilon_{0}}{2}\chi_{\mathrm{ee}}^{\theta \theta}M^{(1)}_{\begin{subarray}{l}e\\o \end{subarray}nm \theta} 
+\frac{j\omega \epsilon_{0}}{2}\chi_{\mathrm{ee}}^{\theta \phi}M^{(1)}_{\begin{subarray}{l}e\\o \end{subarray}nm \phi} 
\right] C_{\begin{subarray}{l}e\\o \end{subarray}nm } + 
\\ 
\left[-\frac{j}{\eta_{0}}M^{(1)}_{\begin{subarray}{l}e\\o \end{subarray}nm \phi}+\frac{j\omega \epsilon_{0}}{2}\chi_{\mathrm{ee}}^{\theta \theta}N^{(1)}_{\begin{subarray}{l}e\\o \end{subarray}nm  \theta} 
+\frac{j\omega \epsilon_{0}}{2}\chi_{\mathrm{ee}}^{\theta \phi}N^{(1)}_{\begin{subarray}{l}e\\o \end{subarray}nm  \phi}
\right] D_{\begin{subarray}{l}e\\o \end{subarray}nm } 
\\
= 
H_{\phi}^{\mathrm{inc}} - \frac{j\omega \epsilon_{0}}{2}\chi_{ee}^{\theta\theta}E_{\theta}^{\mathrm{inc}}
- \frac{j\omega \epsilon_{0}}{2}\chi_{ee}^{\theta\phi}E_{\phi}^{\mathrm{inc}}
\end{split}
\end{equation}
\begin{equation}
\begin{split}
\sum_{n=1}^{N} \sum_{m=0}^{n} 
\\ 
\left[M^{(4)}_{\begin{subarray}{l}e\\o \end{subarray}nm \theta} - \frac{\omega \mu_{0}}{2 \eta_{0}}\chi_{\mathrm{mm}}^{\phi \phi}N^{(4)}_{\begin{subarray}{l}e\\o \end{subarray}nm \phi} - \frac{\omega \mu_{0}}{2 \eta_{0}}\chi_{\mathrm{mm}}^{\phi \theta}N^{(4)}_{\begin{subarray}{l}e\\o \end{subarray}nm \theta} \right] A_{\begin{subarray}{l}e\\o \end{subarray}nm} +
 \\
 \left[N^{(4)}_{\begin{subarray}{l}e\\o \end{subarray}nm \theta} - \frac{\omega \mu_{0}}{2\eta_{0}}\chi_{\mathrm{mm}}^{\phi\phi}M^{(4)}_{\begin{subarray}{l}e\\o \end{subarray}nm\phi} 
- \frac{\omega \mu_{0}}{2\eta_{0}}\chi_{\mathrm{mm}}^{\phi\theta}M^{(4)}_{\begin{subarray}{l}e\\o \end{subarray}nm\theta} \right] B_{\begin{subarray}{l}e\\o \end{subarray}nm} +
\\
\left[-M^{(1)}_{\begin{subarray}{l}e\\o \end{subarray}nm \theta} -\frac{\omega \mu_{0}}{2 \eta_{0}}\chi_{\mathrm{mm}}^{\phi \phi}N^{(1)}_{\begin{subarray}{l}e\\o \end{subarray}nm \phi} - \frac{\omega \mu_{0}}{2 \eta_{0}}\chi_{\mathrm{mm}}^{\phi \theta}N^{(1)}_{\begin{subarray}{l}e\\o \end{subarray}nm \theta}\right] C_{\begin{subarray}{l}e\\o \end{subarray}nm} +
\\
\left[-N^{(1)}_{\begin{subarray}{l}e\\o \end{subarray}nm \theta} - \frac{\omega \mu_{0}}{2 \eta_{0}}\chi_{\mathrm{mm}}^{\phi \phi}M^{(1)}_{\begin{subarray}{l}e\\o \end{subarray}nm\phi}  - \frac{\omega \mu_{0}}{2 \eta_{0}}\chi_{\mathrm{mm}}^{\phi \theta}M^{(1)}_{\begin{subarray}{l}e\\o \end{subarray}nm\theta}\right] D_{\begin{subarray}{l}e\\o \end{subarray}nm} 
\\
= 
E_{\theta}^{\mathrm{inc}} - \frac{j\omega \mu_{0}}{2}\chi_{\mathrm{mm}}^{\phi\phi}H_{\phi}^{\mathrm{inc}}
- \frac{j\omega \mu_{0}}{2}\chi_{\mathrm{mm}}^{\phi\theta}H_{\theta}^{\mathrm{inc}}
\end{split}
\end{equation}
\begin{equation}
\begin{split}
\sum_{n=1}^{N} \sum_{m=0}^{n}  \\
\left[M^{(4)}_{\begin{subarray}{l}e\\o \end{subarray}nm \phi}+\frac{\omega \mu_{0}}{2 \eta_{0}}\chi_{\mathrm{mm}}^{\theta\theta}N^{(4)}_{\begin{subarray}{l}e\\o \end{subarray}nm \theta} +\frac{\omega \mu_{0}}{2 \eta_{0}}\chi_{\mathrm{mm}}^{\theta\phi}N^{(4)}_{\begin{subarray}{l}e\\o \end{subarray}nm \phi}\right] A_{\begin{subarray}{l}e\\o \end{subarray}nm} + 
 \\
\left[N^{(4)}_{\begin{subarray}{l}e\\o \end{subarray}nm\phi}+\frac{\omega \mu_{0}}{2\eta_{0}}\chi_{\mathrm{mm}}^{\theta\theta}M^{(4)}_{\begin{subarray}{l}e\\o \end{subarray}nm\theta} +\frac{\omega \mu_{0}}{2\eta_{0}}\chi_{\mathrm{mm}}^{\theta\phi}M^{(4)}_{\begin{subarray}{l}e\\o \end{subarray}nm\phi}\right] B_{\begin{subarray}{l}e\\o \end{subarray}nm} + 
\\
\left[-M^{(1)}_{\begin{subarray}{l}e\\o \end{subarray}nm \phi}+\frac{\omega \mu_{0}}{2 \eta_{0}}\chi_{\mathrm{mm}}^{\theta \theta}N^{(1)}_{\begin{subarray}{l}e\\o \end{subarray}nm \theta} +\frac{\omega \mu_{0}}{2 \eta_{0}}\chi_{\mathrm{mm}}^{\theta \phi}N^{(1)}_{\begin{subarray}{l}e\\o \end{subarray}nm \phi} \right] C_{\begin{subarray}{l}e\\o \end{subarray}nm} + 
\\
\left[-N^{(1)}_{\begin{subarray}{l}e\\o \end{subarray}nm \phi}+\frac{\omega \mu_{0}}{2 \eta_{0}}\chi_{\mathrm{mm}}^{\theta \theta}M^{(1)}_{\begin{subarray}{l}e\\o \end{subarray}nm\theta} +\frac{\omega \mu_{0}}{2 \eta_{0}}\chi_{\mathrm{mm}}^{\theta \phi}M^{(1)}_{\begin{subarray}{l}e\\o \end{subarray}nm\phi} \right] D_{\begin{subarray}{l}e\\o \end{subarray}nm}
\\
= 
E_{\phi}^{\mathrm{inc}} + \frac{j\omega \mu_{0}}{2}\chi_{\mathrm{mm}}^{\theta\theta}H_{\theta}^{\mathrm{inc}} + \frac{j\omega \mu_{0}}{2}\chi_{\mathrm{mm}}^{\theta\phi}H_{\phi}^{\mathrm{inc}}
\end{split}
\end{equation}
In the equations, the spatial  dependance of the VWF components $(a,\theta,\phi)$, susceptibility tensor components ($\theta,\phi$) and incident field $(a,\theta,\phi)$ are suppressed for brevity. The linear equations are solved by using point matching on the metasurface sphere. For a truncation order $N$, the total number of unknowns is $8 \times \frac{N}{2}\left(N+3\right)$. Let $N_{\theta},N_{\phi}$ be the number of sampling points along the $\theta,\phi$ directions. Then the total number of equations is $4N_{\theta} N_{\phi}$. The metasurface sphere is sampled densely so that $4N_{\theta} N_{\phi} \gg 8 \times \frac{N}{2}\left(N+3\right)$. This results in an overdetermined system of equations which are solved by least squares method.

\section{Numerical validation}
This section contains two illustrative examples validating the proposed method. For both the cases the VWF expansion is truncated to $N = 8$. The metasurface sphere radius is $4\lambda$  and it is sampled with 1600 points for point matching. The simulation duration is around few minutes in MATLAB computing environment on a standard personal computer platform.
\subsection{Illusion transformation}
\begin{figure}[!h]
\centering
\includegraphics[width=3.7in]{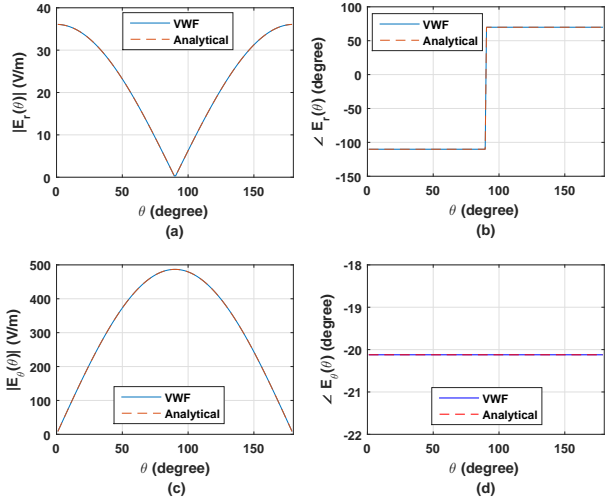}
\caption{Illusion transformation: electric field external to metasurface. (a). $|E_{r}(4.3\lambda,\theta,60^{0})|$ (b).
$\protect \angle E_{r}(4.3\lambda,\theta,60^{0})$ (c).  $|E_{\theta}(4.3\lambda,\theta,60^{0})|$ (d).
 $\protect \angle E_{\theta}(4.3\lambda,\theta,60^{0})$
}
\label{fig_EFieldExtComp}
\end{figure}

\begin{figure}[!h]
\centering
\includegraphics[width=3.7in]{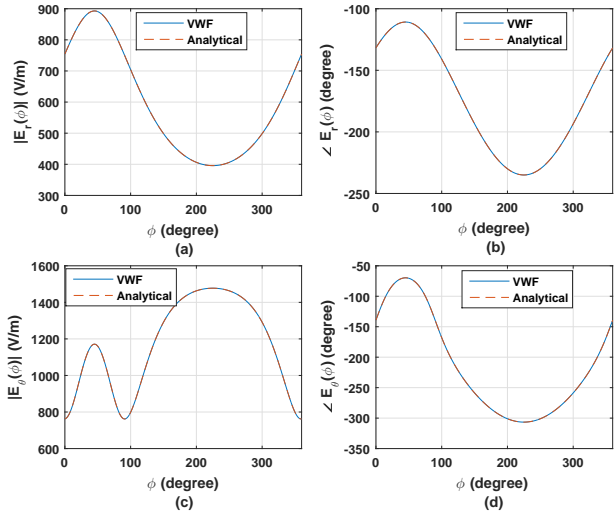}
\caption{Illusion transformation: electric field internal to metasurface. (a) $|E_{r}(0.5\lambda,135^{0},\phi)|$ (b) $\protect\angle E_{r}(0.5\lambda,135^{0},\phi)$ (c) $|E_{\theta}(0.5\lambda,135^{0},\phi)|$ (d) $\protect\angle E_{\theta}(0.5\lambda,135^{0},\phi)$  }
\label{fig_EFieldIntComp}
\end{figure}

The proposed method is numerically validated by a case of illusion transformation as explained in \cite{SphericalMS}. Consider a metasurface that can transform the field generated by an off-centered electric impulse current density, $\bar{J}_{\!s}(\bar{r}^{'}) = \delta(x^{'}-x_{s}) \delta(y^{'}-y_{s}) \delta(z^{'}-z_{s}) \hat{z}$ to field generated by an electric impulse current density at origin, $\bar{J}_{\!s}(\bar{r}^{'}) = \delta(x^{'}) \delta(y^{'}) \delta(z^{'}) \hat{z}$. The location of the displaced source is $x_{s} = \lambda/3,y_{s} = \lambda/3,z_{s} = \lambda/3$. The metasurface is reflection less (i.e. the field inside the metasurface is just that of displaced electric impulse). Such a transformation can be achieved by a monoanisotropic, non-gyrotropic metasurface. Non-gyrotropic metasurfaces are defined by $\chi_{\mathrm{ee}}^{\theta \phi} =
\chi_{\mathrm{ee}}^{\phi \theta} = \chi_{\mathrm{ee}}^{\theta \phi} =
\chi_{\mathrm{ee}}^{\phi \theta} = 0$. The synthesis procedure is explained in detail in \cite{SphericalMS}. In Fig. \ref{fig_EFieldExtComp}, field external to the metasurface is compared with analytical results. Similary, the internal field is compared with analytical results in Fig. \ref{fig_EFieldIntComp}. Excellent agreement can be seen between the proposed method and expected results.

\subsection{Birefringent metasurface}
\begin{figure}[!h]
\centering
\includegraphics[width=3.7in]{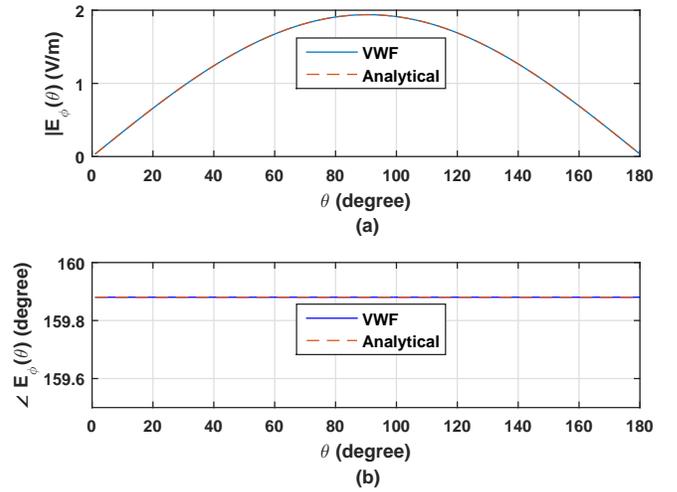}
\caption{Birefringent metasurface: electric field external to metasurface. (a) $|E_{\theta}(4.3\lambda,\theta,60^{0})|$ (b) $\protect\angle E_{\theta}(4.3\lambda,\theta,60^{0})$ }
\label{fig_EFieldCompBirefr}
\end{figure}

A birefringent metasurface is capable of performing two independent transformations. The two transformations for this example are: Transformation\ 1:
field generated by an impulse electric source is transformed to field generated by an impulse magnetic source. Transformation\ 2: field generated by an impulse magnetic source is attenuated by a factor of 2 (both electric and magnetic fields). Since there are two transformations the metasurface will be gyrotropic. The metasurface is illuminated simultaneously by electric and magnetic impulse source. So the expected field external to the metasurface is $1.5\ \times $ the field due to a magnetic impulse source. Excellent agreement between the simulation and expected results can be seen in Fig. \ref{fig_EFieldCompBirefr}.

\section{Conclusion}
A novel approach based on vector wave function expansion combined with bianisotropic susceptibility tensor model is presented for fast and accurate analysis of spherical metasurfaces. The method allows for the determination of field both internal and external to the metasurface. Even though, only source internal to the metasurface is considered in this letter, the method can be easily extended to the case where sources are present both internal and external to the metasurface. The method can probably be used for closed, non-spherical metasurfaces.


%




\ifCLASSOPTIONcaptionsoff
  \newpage
\fi



\bibliographystyle{IEEEtran}
\bibliography{IEEEabrv,BibDataBase}

\vfill

\end{document}